\documentclass[10pt]{article}
\usepackage{amsmath}
\usepackage{amssymb}
\usepackage{graphicx}

\usepackage{cite}

\usepackage{color} 

\usepackage[labelfont=bf,labelsep=period,justification=raggedright]{caption}

\makeatletter
\renewcommand{\@biblabel}[1]{\quad#1.}
\makeatother

\date{}

\usepackage{threeparttable}

\begin{document}

\begin{flushleft}
{\Large
\textbf{Modeling Regulation of Zinc Uptake via ZIP Transporters in Yeast and Plant Roots}
}
\\
Juliane Claus$^{1}$, 
 Andr\'es Chavarr\'ia-Krauser$^{1*}$
\\
\bf{1} Center for Modelling and Simulation in the Biosciences (BIOMS),
	 Universit\"at Heidelberg,
	   INF 368,
	   69120 Heidelberg, Germany
\\
$\ast$ E-mail: andres.chavarria@bioquant.uni-heidelberg.de
\end{flushleft}

\section*{Abstract}

In yeast (\textit{Saccharomyces cerevisiae}) and plant roots (\textit{Arabidopsis thaliana}) zinc enters the cells via influx transporters of the ZIP family. Since zinc is both essential for cell function and toxic at high concentrations, tight regulation is essential for cell viability.  We provide new insight into the underlying mechanisms, starting from a general model based on ordinary differential equations and adapting it to the specific cases of yeast and plant root cells. In yeast, zinc is transported by the transporters ZRT1 and ZRT2, which are both regulated by the zinc-responsive transcription factor ZAP1. Using biological data, parameters were estimated and analyzed, confirming the different affinities of ZRT1 and ZRT2 reported in the literature. Furthermore, our model suggests that the positive feedback in ZAP1 production has a stabilizing function at high influx rates. In plant roots, various ZIP transporters are involved in zinc uptake. Their regulation is largely unknown, but bZIP transcription factors are thought to be involved. We set up three putative models: activator only, activator with dimerization and activator/inhibitor. These were fitted to measurements and analyzed. Simulations show that the activator/inhibitor model outperforms the other two in providing robust and stable homeostasis at reasonable parameter ranges.

\section*{Author Summary}

The heavy metal zinc is both an essential micronutrient for all living organisms and a toxin at high concentrations. Zinc uptake in plants has become an especially interesting research area, because certain hyperaccumulating plant species can be used to clean up zinc-contaminated soils in mining or industrial areas. On the other hand, crops may be manipulated to accumulate zinc in order to overcome nutritional zinc deficiency, which is a problem in many developing countries. Since plants and fungi rely on the zinc content of the surrounding medium, they need to have mechanisms providing a tight uptake regulation at varying external zinc concentrations. In view of the promising applications it is important to understand these mechanisms. Here, we want to support experimental findings with some purely theoretical insights. We use quantitative data from plants and yeast to develop feasible models, analyze their properties in a general framework and discuss their biological relevance. The results confirm experimental hypotheses and may give input for further measurements.

\section*{Introduction}

Zinc is a heavy metal and micronutrient that plays an important role in all living organisms and is particularly essential for the growth of higher green plants \cite{Sommer_1926}.  It is part of the functional subunits or cofactor of more than 300 proteins, among them the class of zinc-finger-proteins as well as RNA-polymerases. In addition, it has been reported to protect plant cells from oxidative stress mediated by reactive oxygen species (ROS) \cite{Cakmak_2000} and may act as an intracellular second messenger \cite{Yamasaki_2007}.

In higher doses, however, zinc becomes toxic. Toxicity is far less frequent than deficiency, but likely in plants growing on contaminated soils, e.g. in mining or industrial areas. Most plants react to elevated zinc levels with toxicity syndromes, such as reduced growth and leaf chlorosis \cite{Broadley_2007}. Only specialized zinc-hyperaccumulating species are able to tolerate high levels without impairment \cite{Zhao_2000}. In order to do so, they possess mechanisms for both the increased uptake of zinc from the soil and its sequestration and detoxification \cite{Macnair_1999}. These mechanisms are subject of ongoing research, as they implicate interesting applications in phytoremediation or nutritional enhancement \cite{Chaney_1997}.

Avoiding both deficiency and toxicity, plants need to take up their required amounts of zinc. Unlike animals they cannot adapt their nutrition accordingly, but depend on the zinc content of the soil. This content may vary considerably in different locations and under different conditions. How are plants able to adapt to this variety?

Charged zinc ions are unable to cross cell membranes freely \cite{Cell}. Instead, they are taken up by specialized transporter proteins. To provide a sufficient intracellular zinc concentration without reaching toxicity, these transporters need to be tightly regulated. The regulatory mechanism has to consist of two parts: sensing of the intracellular zinc concentration and reaction to changes by controlling the amounts of zinc transporters.

The sensing of changes in zinc concentrations must be very sensitive, because the actual internal zinc concentration is believed to be very small. Zinc ions bind to various intracellular proteins, are chelated and sequestered into specific cellular compartments, such as the vacuole \cite{Clemens_2001}. Therefore, although the zinc content in the entire cells may be in a millimolar range, the actual concentration of free zinc ions is estimated to be in a femtomolar range \cite{Outten_2001}. Zinc influx carriers are thought to be regulated by this pool of free zinc ions plus ions that are loosely bound to chelator proteins and can be set free to bind to other proteins with higher affinity. 

\subsection*{Models of homeostasis}\label{sec:model_homeostasis}

Homeostatic regulation in biological systems is based on genetic regulatory systems, and ultimately, on concentrations. These are positive, which constrains the possibilities of control substantially. \cite{Ni_2009} showed that the positiveness constraint of a perfectly / robustly regulating enzyme leads to the need of two separate control mechanisms: for influx and efflux. The homeostatic model proposed by \cite{Ni_2009} is
\begin{eqnarray}\label{eq:homeostasis_general}
 \frac{d S}{d t} & = & I - E \ , \\
 \frac{d R}{d t} & = & k\, (S-S_s) \nonumber\ ,
\end{eqnarray}
where $S$ is the regulated species, $I=I(S,R) \geq 0$ and $E=E(S,R) \geq 0$ are the influx and efflux, respectively, $R$ is the regulator, $k$ is a coefficient (not necessarily positive) and $S_s$ is the set point concentration. The above model results in non-physical negative concentrations of the regulator \cite{Ni_2009}. Independently of the type of mechanism sought after, the negative term in $d R/ d t$ needs certain properties to achieve robustness based on positive concentrations. The approach is to have a term which is linear in $R$ for small $R$ (positiveness), but becomes almost independent of $R$ for larger $R$ (robustness) \cite{Ni_2009}.

\begin{figure}[!ht]
\begin{center}
\includegraphics[width=3.27in]{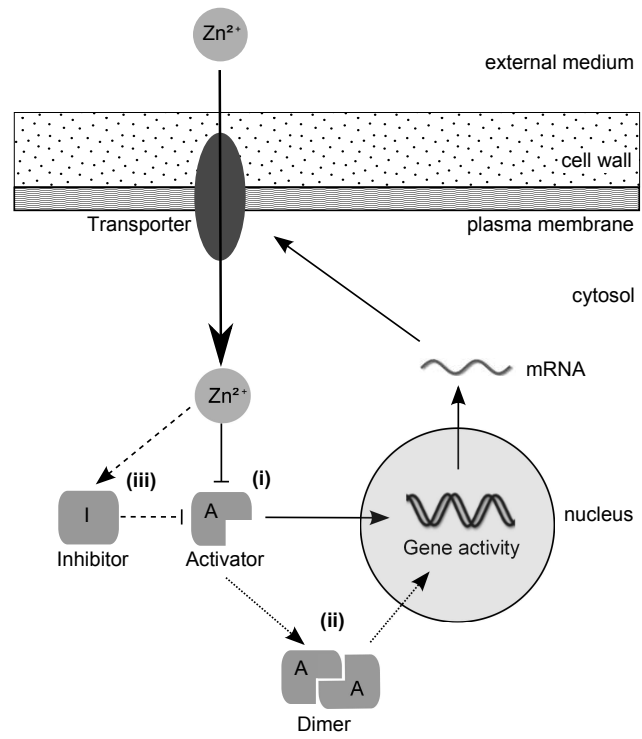}
\end{center}
\caption{
{\bf Plant roots: Scheme of the three models of zinc uptake regulation.}  (i) Activator only, (ii) Activator with dimerization, (iii) Activator/Inhibitor model.
}
\label{fig:Alles}
\end{figure}

Eq.\ \eqref{eq:homeostasis_general} is an oversimplification of homeostatic control in cells, as substantially more complex mechanisms are needed (compare Figs. \ref{fig:ZRT_model} and \ref{fig:Alles}). Also the concept of perfect control is an idealization. Control of zinc fails in cells for low and high external conditions. The presence of oscillations in perfect homeostasis, \cite{Jolma_2010}, poses a problem to living organisms. Strong oscillations could lead to transient, very high and potentially lethal concentrations. Prescinding from perfect regulation could be a compromise between avoiding strong bursts and achieving good control.\\

Based on biological information available, we will develop several putative models of influx homeostasis in plant root cells. In the first part, a general influx regulation model based on an ordinary differential equation system describing transporter gene expression, will be developed and non-dimensionalized. Using the general model, the biological model for yeast of \cite{Zhao_1998} will be translated into a corresponding mathematical model. This model is simplified and fitted to transcript level data via a non-linear optimization method \cite{Gegenfurtner_1992}. The mathematical properties of the steady state, such as stability, are analyzed and discussed. In the second part, the experiences won with the yeast model are used to pose three models for plant roots. The possibilities are manifold, for which reason we restrict the models to the most simple cases of: activator only, activator with dimerization and activator-inhibitor. 

\section*{Results and Discussion}

\subsection*{General model}\label{sec:general_model}

The zinc homeostasis mechanisms presented in this manuscript can be arranged into a general model, which will be developed in this section. Zinc homeostasis can be split into two components: short and long term regulation. Short term regulation is fast but rough, while fine tuning is done by long term regulation. The time scale of short term regulation is less than two hours in plant roots \cite{Talke_2006}. Long term regulation has a substantially larger time scale of several hours, days, weeks, etc.\\

We are interested here in short term regulation, which is local in the sense that the processes occur at the level of single cells in plant roots. Signals besides the fluxes seem not to be transmitted between cells or tissues. This is of course not the case for long term homeostatic control, which will rely definitely on signals transmitted from tissue to tissue. The short term response in plant roots and yeast cells is expected to follow similar laws and can be subdivided into the phases
\begin{equation}\label{eq:general_concept}
\textrm{sensing} \longrightarrow \textrm{transduction} \longrightarrow \textrm{reaction}
\end{equation}
The zinc status is measured in the sensing phase, decisions are taken in the transduction phase and changes in cytosolic concentration occur in the reaction phase. As mentioned in Section \ref{sec:model_homeostasis}, both influx and efflux can be adapted to achieve homeostatic control. In plant roots as well as in yeast cells, adaptation of the expression of influx transporters poses the major component of zinc regulation \cite{Eide_2003,Talke_2006}.

Based on the concept presented in Eq.\ \eqref{eq:general_concept}, the models considered in this manuscript have the following structure
{\small
\begin{equation}\label{eq:general_model}
\begin{array}{p{0mm}l}
\multicolumn{2}{l}{\textit{Sensing:}}\\
& \begin{array}{lcll}
 \frac{d A_i}{d t} & = & p_{Ai}(A_i,\ldots)\quad\!\  - \left(\sum\limits_{j=1}^{n_I} \beta_{ij}\, I_j\ \fbox{$+ \beta_{Ai}\, Z$} + \gamma_{Ai}\right)\, A_i & \ \qquad , \ i=1,\ldots, n_A \ ,\\[3mm]
 \frac{d I_i}{d t} & = & \fbox{$p_{Ii}(I_i,Z, \ldots)$} - \left(\sum\limits_{j=1}^{n_A} \beta_{ij}\, A_j \ \fbox{$+ \beta_{Ii}\, Z$} + \gamma_{Ii}\right)\, I_i & \ \qquad ,\ i=1,\ldots, n_I \ ,\\[3mm]
 \frac{d T_i}{d t} & = &  \alpha_{Ti}\, M_i - \gamma_{Ti}\, T_i \ \fbox{$- \beta_{Ti}\, T_i\, Z$} & \ \qquad ,\ i=1,\ldots, n_T \ ,\\[3mm]
 \end{array}\\[5mm]
\multicolumn{2}{l}{\textit{Transduction:}}\\
& \begin{array}{lcll}
 \frac{d G_i}{d t} & = & \fbox{$\tilde{\mathcal{A}}_i\, \big((1+\tilde{\mathcal{I}}_i)^{-1}-G_i\big)$} - \gamma_{Gi}\, G_i & \\[2mm]
 \frac{d M_i}{d t} & = & \fbox{$\alpha_{Mi}\, G_i$} - \gamma_{Mi}\, M_i & \quad\qquad\qquad\qquad\qquad \ , \ i=1,\ldots, n_T \ ,\\[2mm]
 \frac{d T_i}{d t} & = & \fbox{$\alpha_{Ti}\, M_i$} - \gamma_{Ti}\, T_i - \beta_{Ti}\, T_i\, Z & \\[5mm]
 \end{array}\\
\multicolumn{2}{l}{\textit{Reaction:}}\\
& \begin{array}{rcl}
 \frac{d Z}{d t} & = & \fbox{$\sum\limits_{j=1}^{n_T} \alpha_j\, T_j\, f(Z^e,K_{j}^t)$} - \sum\limits_{j=1}^{n_E} \beta_j E_j\, f(Z,K_{j}^e) - \gamma\, Z \ ,
\end{array}
\end{array}
\end{equation}}%
where $Z$ and $Z^e$ are the cytosolic and external zinc concentrations, respectively, $A_i$ are activators, $I_i$ inhibitors, $T_i$ and $E_i$ influx and efflux transporters, respectively,  $G_i$ and $M_i$ the levels of gene expression and mRNA of $T_i$, respectively, and $p_{Ai}$ and $p_{Ii}$ are model dependent production terms. The total activation and repression are
\begin{equation}\label{eq:act_inhib_dim}
 \tilde{\mathcal{A}}_i = \sum\limits_{j=1}^{n_A} \alpha_{ij}\, A_j\, + \sum\limits_{j,k=1}^{n_A}\alpha_{ij}^{\ k}\, A_j\, A_k\ \quad \textrm{and}\quad  \tilde{\mathcal{I}}_i = \sum\limits_{j=1}^{n_I} \kappa_{ij} I_j \ ,
\end{equation}
and the function $f(Z,K)$ describes saturation of the transporters
\[
 f(Z,K) = \frac{Z}{Z + K} \ .
\]

Sensing is assumed to take place at the level of the activators $A_i$ and inhibitors $I_i$. The possibility that the transporters $T_i$ sense the cytosolic zinc concentration $Z$ directly was also introduced. To achieve regulation, the total activation $\tilde{\mathcal{A}}_i$ has to decrease with higher $Z$ values (see Section \ref{sec:model_homeostasis}). Transduction is modeled in the usual way \cite{Keener}. Three equations per protein are needed, namely for: gene activity $G_i$, transcription into $M_i$ and translation into $T_i$. The activators are introduced as essential transcription factors, which activate the gene transcription. The quadratic form in Eq.\ \eqref{eq:act_inhib_dim} allows to include dimerization. The inhibitors inhibit either the activators or repress through $\tilde{\mathcal{I}}_i$ directly gene activity. Gene repression was assumed to be non-competitive and fast compared to activation, i.e. it is in quasi-equilibrium and $\kappa_{ij}$ are equilibrium constants. The reaction phase is described by an equation for the cytosolic zinc concentration, which contains essentially the difference between influx and efflux mediated by $T_i$ and $E_i$, respectively, and a transporter independent consumption -$\gamma\, Z$. Regulation of the efflux transporters $E_i$ was left out of Eq.\ \eqref{eq:general_model}, as these vary only slightly in roots and no information on yeast was available. In essence, these proteins would follow a similar transduction system as the influx transporters $T_i$.\\

\noindent
Non-dimensionalization of transduction in Eq.\ \eqref{eq:general_model} is straightforward using
\[
 M_{0,i} = \frac{\alpha_{Mi}}{\gamma_{Mi}} \ , \quad T_{0,i} = \frac{\alpha_{Ti}}{\gamma_{Ti}}\, M_{0,i} \ , \quad \Gamma_{Ti} =  \frac{\beta_{Ti}}{\gamma_{Ti}}\, Z_0 \ ,
\]
and the non-dimensionalized total activation and repression
\begin{equation}\label{eq:total_act_inhib}
 \mathcal{A}_i = \sum\limits_{j=1}^{n_A} K_{ij}\, A_j + \sum_{j,k}^{n_A} K_{ij}^{\ k}\, A_j\,  A_k \quad \textrm{and} \quad \mathcal{I}_i = \sum\limits_{j=1}^{n_I} K_{ij}'\, I_j\ ,
\end{equation}
with
\[
 K_{ij} = \frac{\alpha_{ij}}{\gamma_{Gi}}\, A_{0,j}\ , \quad K_{ij}^{\ k} = \frac{\alpha_{ij}^{\ k}}{\gamma_{Gi}}\, A_{0,j}\, A_{0,k} \ , \quad \textrm{and}\quad K_{ij}' = \frac{\kappa_{ij}}{\gamma_{Gi}}\, I_{0,j}\ .
\]
Reaction is non-dimensionalized by choosing
\[
 Z_0 = \frac{\alpha_1}{\gamma}\, T_{0,1}\ , \quad \kappa_j = \frac{\alpha_j}{\alpha_1}\, \frac{T_{0,j}}{T_{0,1}} \quad \textrm{and}\quad \Gamma_j := \frac{\beta_j}{\gamma} E_{0,j} \ .
\]
Non-dimensionalization of the sensing equations depends on the particular structure of the production terms. The decay terms can be non-dimensionalized choosing
\[
\Gamma_{ij} = \frac{\beta_{ij}}{\gamma_{Ai}}\, I_{0,j} \ ,\qquad \Gamma_{ij}' = \Gamma_{ij}\,\frac{\gamma_{Ai}}{\gamma_{Ii}}\, \frac{A_{0,j}}{I_{0,j}} \ , \qquad \Gamma_{Ai} = \frac{\beta_{Ai}}{\gamma_{Ai}}\, Z_0 \ ,\qquad \Gamma_{Ii} = \frac{\beta_{Ii}}{\gamma_{Ii}}\, Z_0  \ ,
\]
while the productions terms still have to be non-dimensionalized accordingly
\[
\frac{1}{\gamma_{Ai}\,A_{0,i}}\,p_{Ai}(A_i,\ldots) \quad \textrm{and} \quad \frac{1}{\gamma_{Ii}\,I_{0,i}}\, p_{Ii}(I_i,Z, \ldots) \ .
\]

\subsection*{Yeast}\label{sec:yeast}

\begin{figure}[!ht]
\begin{center}
\includegraphics[width=3.27in]{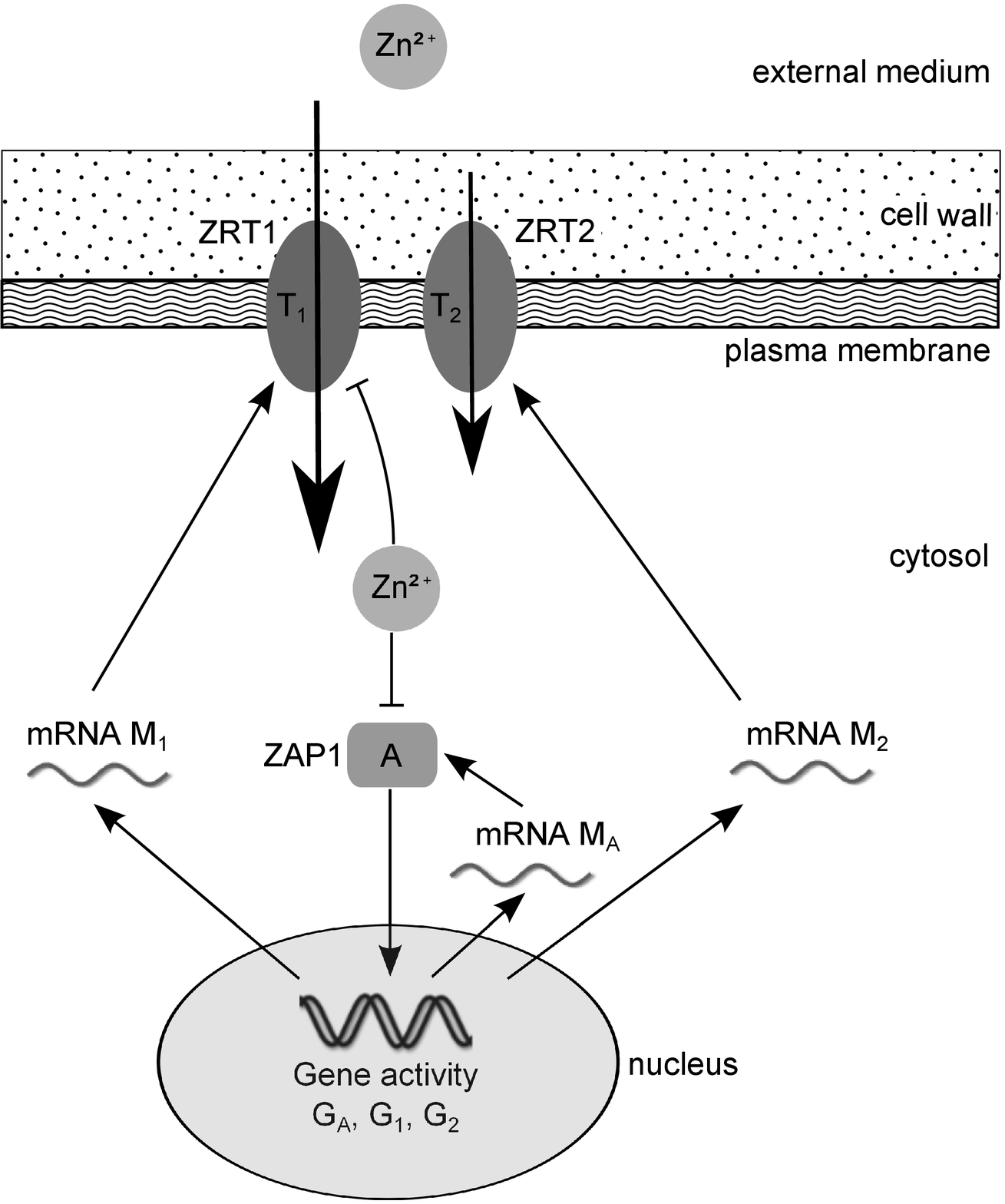}
\end{center}
\caption{
{\bf Yeast: scheme of zinc influx regulation model.} ZAP1 is inactivated by zinc and activates transcription of the transporters ZRT1 and ZRT2.
}
\label{fig:ZRT_model}
\end{figure}

The regulation of zinc uptake in yeast cells (\textit{Saccharomyces cerevisiae}) has been studied in much detail and found to be a combination of two systems with high and low affinity for zinc ions. A similar distribution of high and low affinity transporters has also been found in wheat plants \cite{Hacisalihoglu_2001} and is thought to exist in other plants as well \cite{Guerinot_2000}. A schematic overview of the system can be seen in Fig.\ \ref{fig:ZRT_model}. Zinc ions are transported with high affinity by ZRT1 (zinc-responsive transporter) and with low affinity by ZRT2, which both belong to the ZIP (zinc-, iron-permease) family. ZRT1 has been found to be strongly regulated by the intracellular zinc concentration and almost exclusively active under conditions of zinc deficiency \cite{ZRT1}. ZRT2 has been reported to guarantee a basic zinc uptake level under normal zinc-replete conditions \cite{ZRT2} while being repressed under zinc deficiency \cite{Bird_2004}. Further studies have shown that both \textit{ZRT1} and \textit{ZRT2} are activated by the transcription factor ZAP1 (zinc-dependent activator protein) \cite{Zhao_1998}, which binds to so-called zinc responsive elements (ZREs) in the promoter regions of the respective genes. Under conditions of elevated zinc concentrations, the activity of ZAP1 is reduced and production of ZRT1 and ZRT2 decreases. Inactivation of ZAP1 occurs most likely by direct binding of free zinc ions, although further signaling molecules may also be involved in this process. By binding to its own promoter region, ZAP1 regulates its transcription introducing a positive feedback mechanism and presumably allowing an even stronger response to zinc-limiting conditions, \cite{Eide_2003}. In addition to the transcriptional regulation, ZRT1 is also regulated by a post-trans\-lational mechanism \cite{Eide_2003}. While it is a stable membrane protein under zinc deficient conditions, ZRT1 is ubiquinated and subjected to endocytosis for high intracellular zinc levels. The exact details of this mechanism have been investigated by \cite{Gitan_2003}, but it is yet unknown 
whether zinc ions bind directly to ZRT1 to induce its ubiquitination, or whether other zinc-binding proteins are involved. It has been proposed that the combination of transcriptional and post-translational regulation allows for a very quick response to changing environmental conditions and thus prevents a toxic zinc shock, \cite{Eide_2003}. An overview of the yeast zinc uptake mechanism is presented in Fig.\ \ref{fig:ZRT_model}.

\subsubsection*{Model}

As described above, zinc uptake regulation in yeast comprises the two zinc transporters ZRT1 and ZRT2, as well as the transcription factor ZAP1 as the only activator, which is directly inhibited by zinc ions without an inhibitor. The production of the activator, which corresponds to the term $p_{Ai}(A_i,...)$ in the general model Eq.\ \eqref{eq:general_model}, is a system of sensing, transduction and regulation by itself, because ZAP1 acts as its own transcription factor through a positive feedback loop. While \textit{ZRT1} is simply activated by ZAP1, \textit{ZRT2} is both activated and repressed by the same molecule \cite{Bird_2004}. Therefore, we assume a model with two binding sites of ZAP1 close to the \textit{ZRT2} gene, one activating and one repressing. The total inactivation $\mathcal{I}_i$ (see Eq.\ \eqref{eq:total_act_inhib}) introduces this mechanism into the general model Eq.\ \eqref{eq:general_model}. Here, the inhibitor is equal to the activator and only the \textit{ZRT2} gene is affected: $\mathcal{I}_1 = 0$ and $\mathcal{I}_2 = K_2'\, A$.
 
Following the syntax of the general model and the non-dimensionalization derived in Section \ref{sec:general_model}, we obtain the following system
\begin{equation} \label{eq:full_ZRT}
\begin{array}{lcll}
	\frac{dG_A}{dt} & = & \gamma_{GA} \, \big(K_{A} \, A\, (1-G_A) -  G_A\big) & \\[2mm]
	\frac{dM_A}{dt} & = & \gamma_{MA} \, (G_A -  M_A) & \\[2mm]
	\frac{dA}{dt} & = & \gamma_A \, (M_A - A - \Gamma_A \, Z\, A) & \\[4mm]
	\frac{dG_1}{dt} & = & \gamma_{G1} \, \big(K_1\, A\, (1-G_1) -  G_1\big), & \\[2mm]
	\frac{dG_2}{dt} & = & \gamma_{G2} \, \Big(K_2\, A\,\big( (1+K_2'\, A)^{-1}-G_2\big) - G_2\Big) & \\[2mm]
	\frac{dM_i}{dt} & = & \gamma_{Mi}\, ( G_i - M_i), & \quad i=1,2 \\[2mm]
	\frac{dT_1}{dt} & = & \gamma_{T1}\, (M_1  - T_1 - \Gamma_{T1}\, T_1\, Z) & \\[2mm]
	\frac{dT_2}{dt} & = & \gamma_{T2}\, (M_2   -T_2) & \\[4mm]
	\frac{dZ}{dt} & = & \gamma\, \big(T_1\, f(Z^e,K^t_1) +  \kappa\, T_2\, f(Z^e,K^t_2) - Z\big) & 
\end{array}
\end{equation}
The post-translational regulation of ZRT1 is given by the term $-\gamma_{T1}\,\Gamma_{T1}\,T_1\,Z$. For simplicity the term $-\gamma\, Z$ accounts for all zinc consumption processes. These may include export from the cell through zinc efflux transporters, sequestration into the vacuole and other compartments as well as binding and chelation of zinc by various proteins in the cytoplasm.

The trivial solution (all species zero) is a steady state of Eq.\ \eqref{eq:full_ZRT}. There is at least one non-trivial steady state, which for the activator ZAP1 can be written as a function of the intracellular zinc concentration
\begin{equation}\label{eq:zap_ss}
	\overline{A} = \frac{1}{1+\Gamma_A \, \overline{Z}} - \frac{1}{K_{A}}.
\end{equation}
For $\overline{A}$ not to become negative, this equation poses the condition $K_{A}>1+\Gamma_A \, \overline{Z}$, which implies that for large $\overline{Z}$ the non-trivial and trivial solutions cross. A detailed analysis of this case is presented below. The case of total deficiency brings insight into some of the parameters. We find as expected $\overline{Z} \to 0$ for $Z^{e}\to 0$, which means that
$\overline{A} \to 1 - 1/K_{A}$. From biological point of view, $\overline{A}$ is expected to shoot to a value near to $1$ for total deficiency, which has as a consequence that $K_A \gg 1$. Assuming that $\overline{A} \approx 1$ for $Z^{e}\to 0$, the concentrations of the transporters $T_1$ and $T_2$ behave as
\[
    \overline{T_1} \to \frac{1}{1 + 1/K_{1}} \quad \textrm{and} \quad \overline{T_2}  \to \frac{1}{1 + {1}/{K_2}  + {K_{2}'}/{K_2} + K_{2}'} \quad \textrm{for}\ Z^e \to 0 \ .
\]
High affinity of ZRT1 and low affinity of ZRT2, i.e $T_1 \approx 1$ and $T_2 \approx 0$ for $Z^e \to 0$, are obtained when the conditions $K_1 \gg 1$ and $K_2' + K_2' / K_2 + 1/ K_2 \gg 1$ are fulfilled.  Considering $K_2 \approx K_1 \gg 1$, the second condition is essentially $K_2' \gg 1$. Expression of ZRT2 is maximal for a ZAP1 concentration of $\overline{A} = (K_2\, K_2')^{-1/2}$, while expression of ZRT1 rises monotonically with $\overline{A}$ and reaches its highest value for $Z^e \to 0$. For a given activation $K_2$, repression $K_2'$ has to be large to shift the expression maximum towards low $\overline A$ and high $Z^e$.\\

Using the quantitative data measured by \cite{Zhao_1998} and \cite{Bird_2004}, we estimated the model parameters via optimization. These measurements are stationary, and thus, the system could be reduced into one with the four unknowns $A$, $T_1$, $T_2$ and $Z$. The parameters obtained are listed in Table \ref{tab:parameters_yeast}. These reflect perfectly the above conditions for $K_A$, $K_1$, $K_2$ and $K_2'$. The model reproduces very well the measurements (Fig.\ \ref{fig:ZRT}). 

\subsubsection*{Roles of ZRT1 and ZRT2}

\cite{Zhao_1998} proposed that ZRT1 and ZRT2 play different roles in zinc uptake of yeast cells. While ZRT1 is most active only in zinc-deficient cells, ZRT2 is transiently active also in zinc-replete cells with external zinc concentration around $1000\,\mu M$. This implies that under low external zinc concentrations ZRT1 contributes the most to the overall zinc uptake, while under high external zinc concentration, ZRT2 acts as the major transporter. Such behavior is confirmed by our model.

\begin{figure}[!ht]
\begin{center}
\includegraphics[width=6.83in]{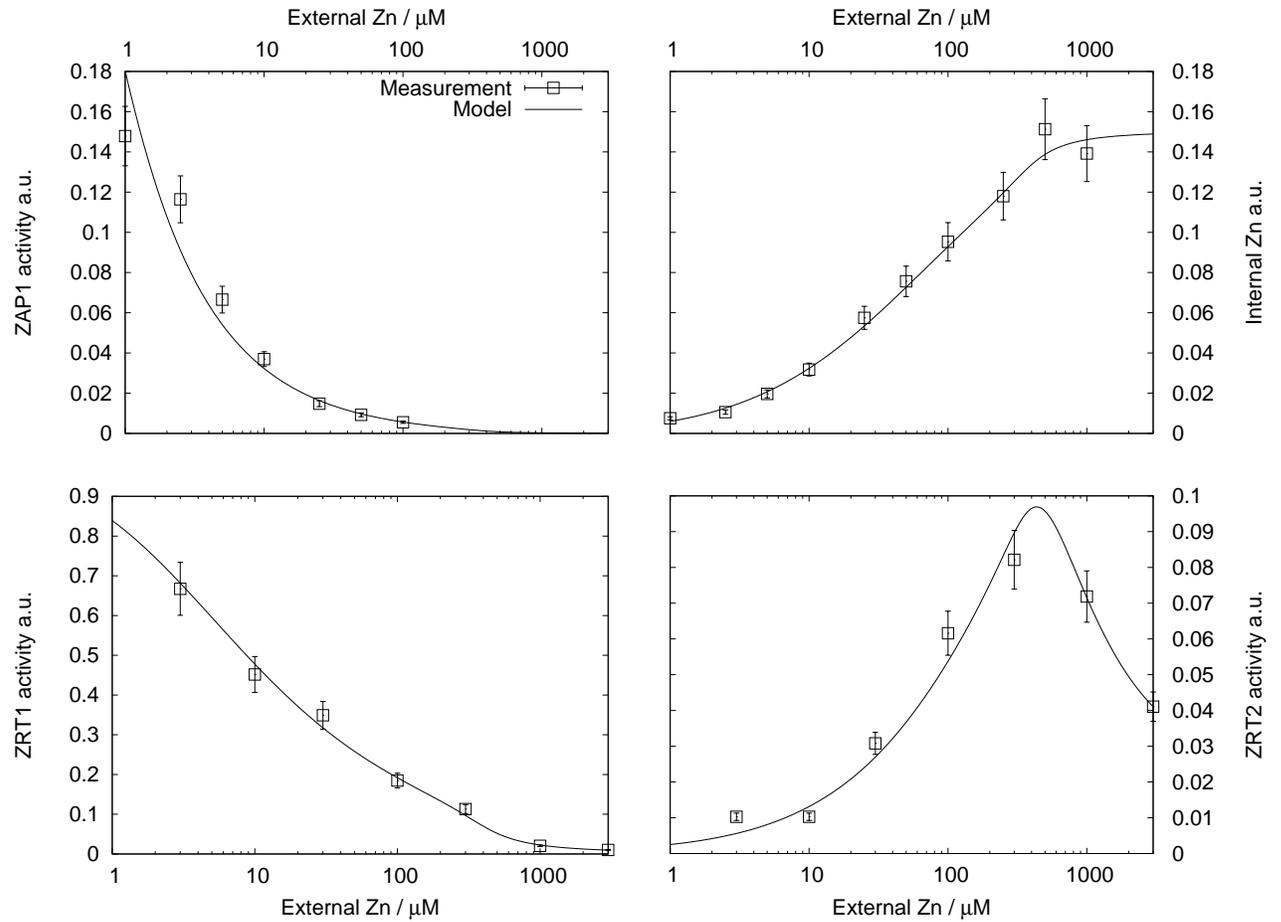}
\end{center}
\caption{
{\bf Yeast simulations:}  Comparison between measurements and simulated steady states of ZAP1, internal zinc, ZRT1 and ZRT2 for varying external zinc concentration. Measurements: ZRT1 and ZRT2 by \cite{Bird_2004}, ZAP1 and zinc by \cite{Zhao_1998}.
}
\label{fig:ZRT}
\end{figure}

\begin{figure}[!ht]
\begin{center}
\includegraphics[width=3.27in]{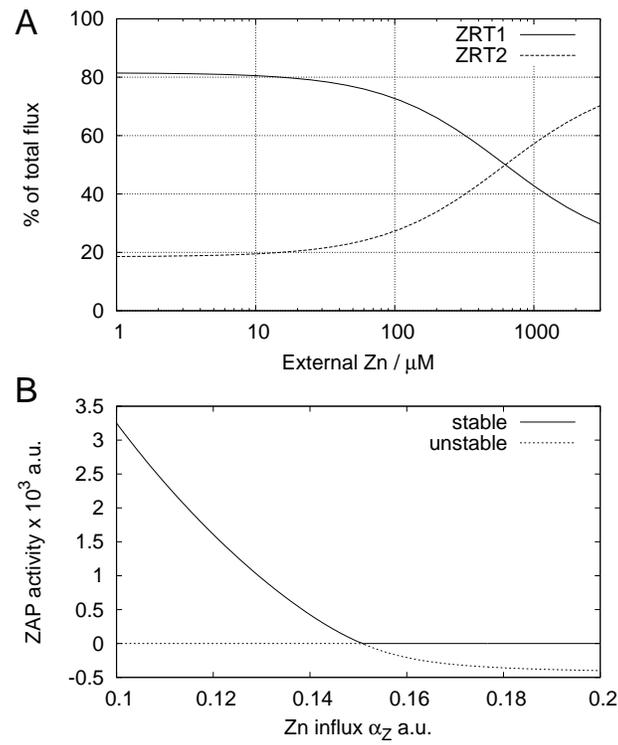}
\end{center}
\caption{
{\bf Yeast: Role of ZRT1 and ZRT2 and ZAP1 feedback.} A, contributions of ZRT1 or ZRT2 to the total zinc influx for varying external zinc concentration. B, ZAP activity for varying values of ZRT independent influx $\alpha_Z$. The stable solution is marked with a solid line, the unstable solution is dotted.
}
\label{fig:bifurcation}
\end{figure}

Fig.\ \ref{fig:bifurcation} \textit{A} presents the relative contributions the the total flux. At  low external concentrations ZRT1 is responsible for about 80\% of flux, while at replete conditions (above 500 $\mu M$) ZRT2 dominates. ZRT1 seems indeed to act as a high affinity transporter with a Michaelis constant $K^t_1 = 139\,\mu M$, while ZRT2 has less affinity reflected by a substantially larger $K^t_2=2584\,\mu M$. A similar ratio was found by \cite{ZRT2}, although their values are several orders of magnitude lower. This discrepancy comes from the assumption made in \cite{ZRT2} that the mechanism is based on Michaelis-Menten kinetics and their low values are reproduced when Michaelis-Menten is fitted to our simulations. The affinity of the ZRT1 and ZRT2 systems are not completely determined by $K^t_1$ and $K^t_2$, respectively. These constants have to be larger than the optimal concentration of the corresponding system, as saturated transporters cannot pass information on external zinc status ($f(Z^e,K^t_i) \approx  1 = \textrm{const}$ for $Z^e \gg K^t_i$). The optimal concentration for \textit{ZRT1} is at total deficiency, while \textit{ZRT2} is most active at $430\,\mu M$ (Fig.\ \ref{fig:ZRT}). 

A strong repression of \textit{ZRT2} is essential to achieve a maximal expression at high external zinc concentrations (see Table \ref{tab:parameters_yeast}). However, a strong repression results also in lower gene activities, which explains why \textit{ZRT2} has a much lower expression level than \textit{ZRT1} (Fig.\ \ref{fig:ZRT} and \cite{Bird_2004}). To counteract the lower expression level, ZRT2 needs either to transport zinc at higher rates or more copies need to be produced. This is reflected by the coefficient $\kappa$, which suggests that ZRT2 is six times more effective in transporting zinc than ZRT1. Assuming that ZRT1 and ZRT2 have similar transport rates, $\kappa \approx 6$ could be a hint for posttranslational regulation of ZRT1, although direct posttranslational regulation via $\Gamma_{T1}$ was shown to be not significant here (F-test: $P>0.05$). 

\textit{ZRT1} and \textit{ZRT2} were found to be activated equally well by ZAP1, as reflected by the small insignificant difference between $K_1$ and $K_2$. The self-activation constant $K_A$ of \textit{ZAP1}, is four times smaller than $K_1$ and $K_2$. This suggests that \textit{ZRT1} and \textit{ZRT2} have four promoters instead of one in \textit{ZAP1}, which is in concord with the experimental results of \cite{Zhao_1998}.

\subsubsection*{ZAP1 transcriptional feedback}\label{sec:zap_feedback}

The feedback loop generated by ZAP1 acting as its own transcription factor brings interesting properties into the model.  \cite{Eide_2003} proposed that this feedback allows a stronger reaction to zinc-limiting conditions. Our model suggests that the advantage is rather for zinc-replete conditions.  The steady state Eq.\ \eqref{eq:zap_ss} of ZAP1 becomes negative for $\overline Z > (K_{A}-1)/\Gamma_A \approx 0.15$ and crosses the trivial steady state. Unless these two steady states exchange their roles, the model would become non-biological at the bifurcation. Based on the fitted parameters, the bifurcation is normally reached at very high external zinc concentrations. To examine the behaviour of the model at the bifurcation, we introduced a ZRT1 and ZRT2 independent path into the cell. The path could for example be another transporter not regulated by ZAP1 and shifts the bifurcation towards lower $Z^e$. Without considering any details of these processes, the simplest modification is to include an additional constant zinc influx term $\alpha_Z$ to the last equation in Eq.\ \eqref{eq:full_ZRT}. The bifurcation is illustrated in Fig.\ \ref{fig:bifurcation} \textit{B}. There are at least two steady states, where one is trivial ($\overline{A}=\overline{T_1}=\overline{T_2}=0$ and $\overline{Z}=\alpha_Z$) and the other is positive for small $\alpha_Z$. The stability of these are exchanged at the bifurcation. For low $\alpha_Z$ the positive steady state is stable, while the trivial steady state is unstable. When the steady states cross at the bifurcation, the trivial solution becomes stable while the now negative steady states becomes unstable. The positive steady state is literally trapped by the trivial steady state. From the biological view the ZAP1 feedback allows the system to completely switch off expression of \textit{ZAP1} and thus of \textit{ZRT1} and \textit{ZRT2}. In a mechanism without feedback, \textit{ZAP1} expression would just fall asymptotically to zero for increasing zinc influx. Therefore, we conclude that the feedback of ZAP1 is advantageous for zinc-replete conditions.

\subsection*{Plant roots}\label{sec:plants}

In plants, zinc is taken up from the soil and transported into the root cells. Unlike in unicellular organisms, zinc needs to be transported into further tissues: xylem, stem, leaves, etc. A number of different transporter proteins are involved. There are three families of transporters for zinc: ZIP, HMA (heavy-metal-ATPases) and MTP (metal tolerance protein) or CDF (cation diffusion facilitator). Members of the ZIP family are believed to act as influx carriers, including  uptake of from the soil (similar to ZRTs in yeast). HMAs accomplish efflux of zinc, e.g. from roots into xylem vessels, while MTPs are involved in sequestration into compartments, such as the vacuole \cite{Palmer_2009}. The main root influx transporters are ZIP1, ZIP2, ZIP3, ZIP9, and IRT3 \cite{Grotz_1998}, while ZIP4 localizes to chloroplast \cite{Guerinot_2000}. These transporters are highly expressed under conditions of zinc deficiency, whereas their expression decreases quickly when zinc is added to the media \cite{Talke_2006}. The exact mechanism of this regulation is still unknown. Recent results showed that at least \textit{ZIP4} in \textit{Arabidopsis thaliana} is regulated by transcription factors of the basic-region leucine zipper (bZIP) family: bZIP19 and bZIP23, \cite{Assuncao_2010}. These factors bind to a ZDRE (zinc deficiency response element), which found in the upstream regions of the \textit{ZIP1}, \textit{ZIP3}, \textit{ZIP4}, and \textit{IRT3}.

Unlike the ZAP1 transcription factor in yeast (see Section \ref{sec:yeast}), bZIP19 and bZIP23 transcription factors do not have a zinc binding site \cite{Assuncao_2010}. It is unclear how these sense the intracellular zinc status. Existence of further players that bind zinc and act as inhibitors of bZIP19 and bZIP23 have been proposed \cite{Assuncao_2010_mini}. Transcription factors of the bZIP family have been studied in other regulatory networks and are known to be regulated post-transcriptionally in various ways \cite{Schutze_2008}. Generally, bZIP transcription factors and in particular bZIP19 and bZIP23 act as dimers \cite{Jakoby_2002}. They are partially redundant \cite{Assuncao_2010} and it is believed that they preferentially form homodimers, but may also interact to constitute heterodimers \cite{Deppmann_2006}.

Our model focuses on the specific transition of zinc into the root cell space. By restricting the model to this specific situation, a similar approach as the one for yeast in Section \ref{sec:yeast} can be applied. We start with a simple model based on one zinc dependent activator. Hereafter, the advantage of dimerization is analysed and a more involved model based on an activator/inhibitor pair is presented. Using the measurements of \cite{Talke_2006}, some of the parameters are obtained via optimization and a F-Test is used to conclude which model is the best. Finally, we analyse the relation between stability and robustness of the activator/inhibitor model.

\subsubsection*{Activator}\label{sec:activatoronly}

Here, we assume that regulation takes place by one zinc dependent transcription factor (see Fig.\ \ref{fig:Alles} \textit{(i)} for a scheme). Based on the general model Eq.\ \eqref{eq:general_model}, we set $n_A=n_T=1$ and $n_I=0$ and avoid unnecessary notation by dropping indexes (e.g. $A=A_1$ and $K=K_{11}$, etc.). Sensing is assumed to take place only at the activator level ($\beta_T = 0$). Also the possibility that the activator acts as a dimer is ruled out ($\alpha_{ij}^{\ k}=0$). Efflux transporters are assumed to be non-saturable, allowing to combine efflux/consumption into one term $-\gamma\, Z$. In contrast to the case of yeast in Section \ref{sec:yeast}, there is no specific information on the production of the activator. To keep the system simple, we introduce a constant pool $A_0$ of activator, which is distributed into active and inactive molecules. The net production is set to $\alpha_A\,(A_0 - A)$ and
\begin{equation}\label{eq:activatoronly_production}
 p_{A} = \alpha_A\, A_0 \qquad \textrm{and}\qquad \alpha_A = \gamma_A\ .
\end{equation}
The non-dimensionalized system is
\begin{eqnarray}\label{eq:activatoronly_model}
 \frac{d A}{d t} & = & \gamma_A\, \big(1 - (1 + \Gamma_A \, Z)\, A \big) \ , \nonumber \\
 \frac{d G}{d t} & = & \gamma_G\, \big(K\,A \, (1-G) - G \big)\ ,\nonumber \\
 \frac{d M}{d t} & = & \gamma_M\, (G - M)\ ,\\
 \frac{d T}{d t} & = & \gamma_T\, (M - T)\ , \nonumber \\
 \frac{d Z}{d t} & = & \gamma \big( T\, f(Z^e,K^t) - Z \big)\ , \nonumber 
 \end{eqnarray}
with two steady states
\begin{eqnarray}\label{eq:activator_ss}
  \overline T & = & \overline M \ = \ \overline G\ =\ \frac{K}{K + 1 + \Gamma_A\, \overline Z}\ , \nonumber \\
  \overline A & = & \frac{1}{1 + \Gamma_A\, \overline Z}\ , \\
  \overline Z & = & \frac{1}{\Gamma_A}\,\left(- \frac{1}{2}\,(K+1) \pm \Big( K\,\Gamma_A\, f(Z^e,K^t) + \frac{1}{4}(K+1)^2 \Big)^{\frac{1}{2}}\right)\ . \nonumber
\end{eqnarray}
The steady state with $\overline Z \leq 0$ is not biologically relevant and is not considered. For total deficiency
\begin{equation}
\overline Z \to  0 \qquad \textrm{and} \qquad \overline G \to \frac{K}{K + 1} \qquad \textrm{for}\ Z^e \to 0 \ .
\end{equation}
Biology suggests that gene expression will shoot to a very high value and $\overline G$ should be near to one: $K \gg 1$. For replete conditions 
\begin{equation}
\begin{array}{rcl}
 \overline Z & \to & \frac{1}{\Gamma_A}\,\left(- \frac{1}{2}\,(K+1) + \Big(K\,\Gamma_A + \frac{1}{4}(K+1)^2 \Big)^{\frac{1}{2}}\right)\\
 \overline G & \to & 1 / \left(\frac{1}{2} \pm \Big( \frac{\Gamma_A}{K} + \frac{1}{4}\Big)^{\frac{1}{2}}\right)\\
\end{array} \qquad \textrm{for}\ Z^e \to \infty\ ,
\end{equation}
where $f(Z^e,K^t) \to 1$ and $K \gg 1$ were used. Biology suggests that gene expression should be small for high external zinc concentrations
\begin{equation}\label{eq:activatoronly_conditions}
\Gamma_A \gg K \gg 1 \ .
\end{equation}
The steady state for varying $Z^e$ depends on three parameters $K$, $\Gamma_A$ and $K^t$. While $K^t$ is a property of the transporters, $K$ and $\Gamma_A$ determine gene activity for extreme conditions. A value $K^t=13\ \mu M$ for ZIP1 was published by \cite{Grotz_1998} and used here. Assuming that gene activity reaches at least 95\% for total zinc deficiency, one obtains
\begin{equation}\label{eq:activatoronly_conditions_vals}
K \geq 20 \ .
\end{equation}
Determination of $K$ from measurements would need data at very low zinc concentrations, which is uncertain and was not available to the authors. For this reason, an empirical value of $K=20$ was used. Only $\Gamma_A$ stays undetermined and a value of $41138$ was obtained by fitting the model to published values of \textit{ZIP3} expression \cite{Talke_2006}. All parameters are listed in Table \ref{tab:parameters_plants}.

Fig.\ \ref{fig:plants_ss} shows the steady state for varying $Z^e$. Gene activity falls slowly for increasing $Z^e$ resulting in a continuously increasing internal zinc concentration. The mechanism breaks for extreme zinc conditions: undersupply at low $Z^e$ and oversupply for large $Z^e$. The reason for oversupply is the activator reacting slowly to changes in $Z^e$. The model offers only a mean to set the extreme gene expressions via $K$ and $\Gamma_A$, and does not allow to set the transition steepness between these. $\Gamma_A$ is also very large compared to the value determined for yeast ($\sim60$ times larger; Table \ref{tab:parameters_yeast}), making the model even more unlikely.

\subsubsection*{Dimerization}\label{sec:dimer}

The transcription factors bZIP19 and bZIP23 are known to act as dimers \cite{Jakoby_2002}. We set $\alpha_{ij}^{\ k} \neq 0$ and $\alpha_{ij} = 0$ in the general model, so that the activator functions only as a dimer. A scheme of the model is presented in Fig.\ \ref{fig:Alles} \textit{(ii)}. The total activation is here $\mathcal{A} = K\, {A}^2$, while the rest stays the same as in Eqs. \eqref{eq:activatoronly_model} and \eqref{eq:activator_ss}, meaning that only gene activity needs to be adapted
\begin{eqnarray}
 \frac{d G}{d t} & = & \gamma_G\, \big(K\,A^2 \, (1-G) - G \big)\ , \\
 \overline{G} & = & \frac{K}{K+(1+\Gamma_A\,\overline{Z})^2}\ . \nonumber
\end{eqnarray}
Because $1 +\Gamma_A\,\overline{Z} \leq (1 +\Gamma_A\,\overline{Z})^2$, gene activity is smaller with dimerization than without (compare Eq.\ \eqref{eq:activator_ss}). Gene activity reacts more sensible to changes of zinc status than in the non-dimersing case (Fig.\ \ref{fig:plants_ss}). The transition between gene on and off is steeper, rendering a more robust mechanism. Fitting the model to the measurements delivers $\Gamma_A = 1844$, which is ca. $20$ times smaller than in the non-dimerizing case and substantially nearer to the value for yeast. From an evolutionary point of view, dimerization allowed to down-regulate the transporters more strongly with less binding affinity. Also, by assuming that the variances of the measured values are proportional to these, one finds that $\chi^2$ is for the model with dimerization less than half as for the one without when fitted to measurements the of \cite{Talke_2006}. In total, the model with dimerization outperforms the model of Sec. \ref{sec:activatoronly}, although these have the same number of degrees of freedom.

\subsubsection*{Activator/Inhibitor} \label{sec:actinhibit}

\begin{figure}[!ht]
\begin{center}
\includegraphics[width=6.83in]{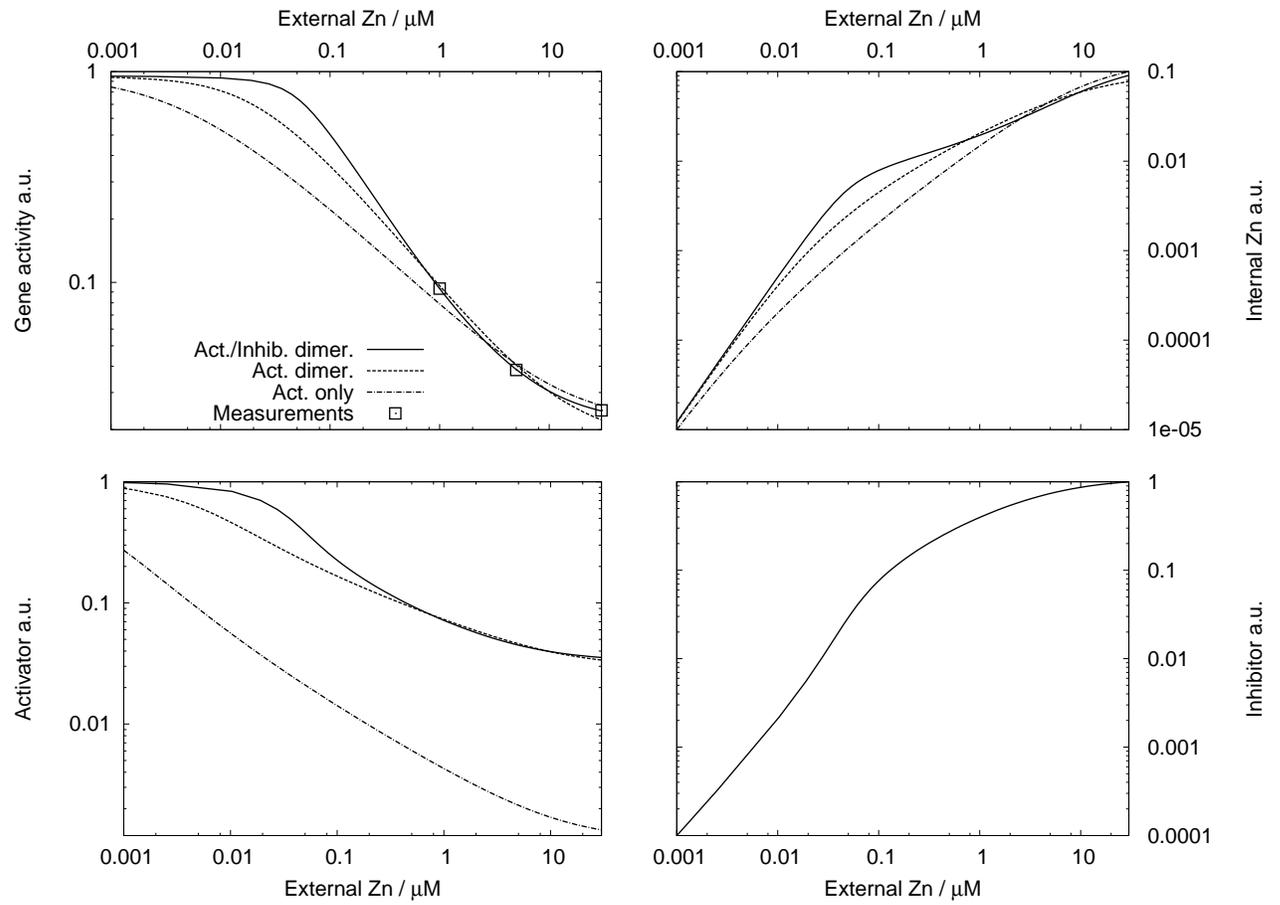}
\end{center}
\caption{
{\bf Plant roots: Steady states of the different regulation models.}  The activator only, dimerizing activator and activator/inhibitor pair with dimerization. Measurements of \cite{Talke_2006} are also shown.
}
\label{fig:plants_ss}
\end{figure}

\begin{figure}[!ht]
\begin{center}
\includegraphics[width=3.27in]{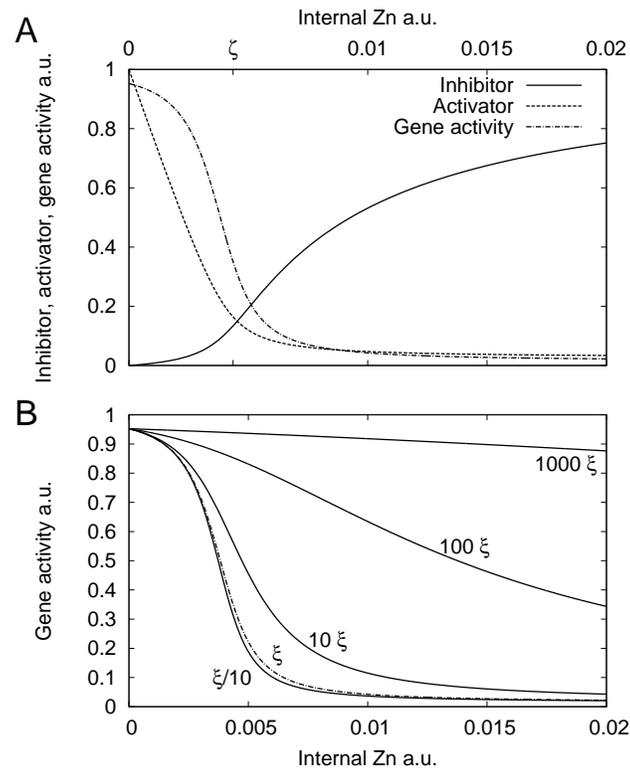}
\end{center}
\caption{
{\bf  Plant roots: Activator/Inhibitor model with dimerization.}  A, steady state values of inhibitor, activator and gene activity in dependence of internal zinc concentration. B, steady state gene activity in dependence of internal zinc status for varying $\xi$. Dashed curve corresponds to the nominal $\xi=10^{-3}$.
}
\label{fig:switch}
\end{figure}

Including dimerization delivered a better fit to the measurements. However, a systematic deviation for higher $Z^e$ was found (Fig.\ \ref{fig:plants_ss}). Following \cite{Assuncao_2010_mini} proposition of intermediate steps in sensing, we propose a mechanism involving an activator/inhibitor pair. Assume that these interact when they are not bound to the DNA, the pairs cannot activate the gene and zinc is sensed only by the inhibitor (Fig.\ \ref{fig:Alles} \textit{(iii)}). Applying this assumptions to the general model Eq.\ \eqref{eq:general_model} delivers $n_A=n_I=n_T=1$. As in Sec. \ref{sec:dimer}, dimerization is included by using the total activation $\mathcal{A} = K\, A^2$. Production of activator is set as in the activator only model (Eq.\ \eqref{eq:activatoronly_production}). Sensing occurs at the level of the inhibitor
\[
 p_I = \alpha_I\, I_0\, Z  \ ,\quad \alpha_I = \beta_I \quad \textrm{and} \quad  \beta_A = 0\ .
\]
Transcription and translation are the same as in the dimerizing activator case. The equation for $Z$ stays the same, meaning that the main differences to Eq.\ \eqref{eq:activatoronly_model} are
\begin{eqnarray}\label{eq:activatorinhibitor_model}
 \frac{d G}{d t} & = & \gamma_G\, \big(K\,A^2 \, (1-G) - G \big)\ ,\nonumber \\
 \frac{d A}{d t} & = & \gamma_A\, \big(1 - \Gamma\, A\, I - A \big) \ , \\
 \frac{d I}{d t} & = & \gamma_I\, \big(\Gamma_I\, Z - \Gamma'\, A\, I - (1 + \Gamma_I \, Z)\, I \big) \ . \nonumber
\end{eqnarray}
If $Z$ is considered to be a parameter in the above system, the steady state is
\begin{eqnarray}
 \overline G & = & \frac{K}{K + (1 + \Gamma\, \overline I )^2} \ , \nonumber \\
 \overline A & = & \frac{1}{1+\Gamma\, \overline I} \ ,\nonumber\\
 \overline I & = & \frac{1}{2}\left(\frac{\overline Z - \zeta}{\overline Z + \xi} -\frac{1}{\Gamma} \right) \pm \left(\frac{1}{\Gamma}\frac{\overline Z}{\overline Z + \xi} + \frac{1}{4}\left(\frac{\overline Z - \zeta}{\overline Z + \xi} -\frac{1}{\Gamma} \right)^2\right)^{\frac{1}{2}} \nonumber \ ,
\end{eqnarray}
where $\zeta = \Gamma'/\Gamma\Gamma_I$ and $\xi = 1/\Gamma_I$. The solution with $\overline I < 0$ is biologically irrelevant. For totally deficient conditions
\[
 \overline I \to 0 \ ,  \quad \overline A  \to 1 \quad \textrm{and} \quad  \overline G \to \frac{K}{K+1} \quad \textrm{for}\ Z^e \to 0\ .
\]
The case of very high external zinc needs to include the expression for $\overline Z$. Instead of determining what happens for $Z^e \to \infty$, we determine the behavior for large internal concentrations
\[
 \overline I \to 1 \ , \quad \overline A \to \frac{1}{1+\Gamma} \quad \textrm{and} \quad  \overline G \to \frac{K}{K+(1+\Gamma)^2} \quad \textrm{for}\ \overline Z \to \infty \ .
\]
The same biological conditions as those listed in Eqs. \eqref{eq:activatoronly_conditions} and \eqref{eq:activatoronly_conditions_vals} are found here. In contrast to the activator models, gene activity does not go to zero for $\overline Z \to \infty$. Again, the constants $\Gamma$ and $K$ determine gene activity for extreme zinc statuses. The steady state values depend on two more constants: $\zeta$ and $\xi$. The first term in $\overline I$ is zero for $\overline Z \approx \zeta$. For $\overline Z < \zeta$ the term is negative and has to be compensated by the slightly larger square root term. Is $\overline Z >\zeta$, then $\overline I (\overline Z)$ grows fast, inhibits the activator and leads to a strong reduction of gene activity (compare Fig.\ \ref{fig:switch} \textit{A}). Thus, $\zeta$ determines the internal zinc concentration for switching the gene from on to off. The constant $\xi$ determines the steepness of the transition between the on and off states (Fig.\ \ref{fig:switch} \textit{B}). A small $\xi$ corresponds to a strong binding affinity $\Gamma_I$ between zinc and inhibitor. The switching steepness is also affected by $\Gamma$, as it weights the first term under the root. Large $\Gamma$ result in steeper switches (effect similar to decreasing $\xi$; Fig.\ \ref{fig:robustness} \textit{B}).

The activator/inhibitor model renders a better and more robust homeostatic control mechanism than the activator only models (Fig.\ \ref{fig:plants_ss}). The reason is the steep genetic switch obtained by the inclusion of an inhibitor, which reacts strongly to the internal zinc status (Fig.\ \ref{fig:switch}). Fitting the model to the measurements delivered $\Gamma=38$ and $\zeta = 4.4 \cdot 10^{-3}$ (Table \ref{tab:parameters_plants}). $\xi$ cannot be determined by a fit, because a robust mechanism is sought after and in that regime the model becomes almost independent of $\xi$ (compare Fig.\ \ref{fig:switch} B). Therefore, a value of the same order as $\Gamma_A$ for yeast was used ($\xi =10^{-3} \Rightarrow \Gamma_I = 1000$ and $\Gamma_A = 714$ for yeast). The model describes the measurements very well (Fig.\ \ref{fig:plants_ss}), which is also a consequence of the small number of degrees of freedom. No systematic deviation for large $Z^e$ was found for this model. An F-Test showed that the activator/inhibitor model performs better irrespective of having one parameter more ($P<0.05$).

\subsubsection*{Robustness and instability} \label{sec:stability}

\cite{Jolma_2010} showed that a perfect homeostatic control can lead to undamped oscillations. In the case of a toxic compound, oscillations could imply lethal peaks. Therefore, the stability of the activator/inhibitor model was analysed. Dynamics and stability depend on the time scales involved in the mechanism. The authors could not find published values for these. Similar values to those listed in \cite{Cook_1998} were used, where the products were assumed to decay four times slower than gene activity. The reader should keep in mind that the specific choice of the time scales influences stability, but the relation between robustness and instability found below should keep its validity.

\begin{figure}[!ht]
\begin{center}
\includegraphics[width=3.27in]{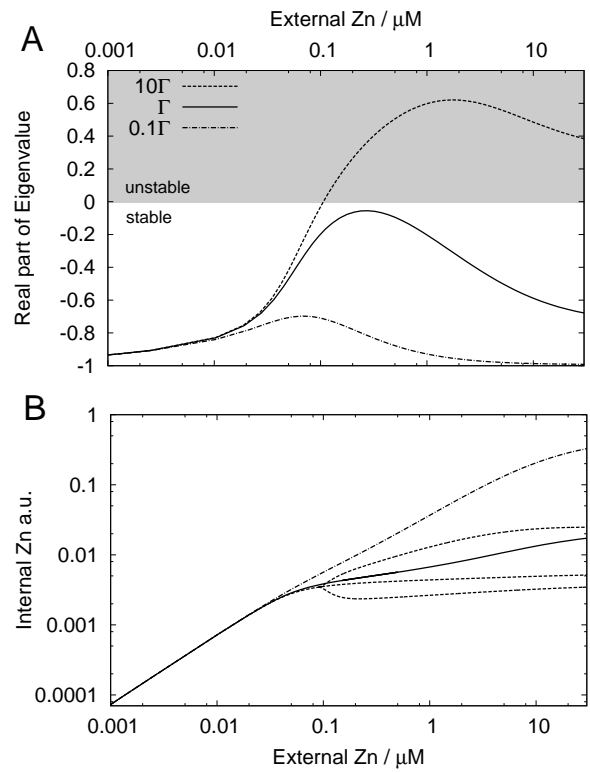}
\end{center}
\caption{
{\bf Plant roots: Robustness and stability.}  Robustness and stability of the activator/inhibitor model for $0.1\Gamma$, $\Gamma$ and $10\Gamma$ and varying external zinc concentration. \textit{A}, real part of largest eigenvalue. \textit{B}, internal zinc concentration. Minimal and maximal values of limit cycle shown for unstable steady state ($10\Gamma$).
}
\label{fig:robustness}
\end{figure}

$\Gamma$ sets the robustness of the activator/inhibitor mechanism by increasing the steepness of the genetic switch. This leads also to instability of the steady state and to undamped oscillations (Fig.\ \ref{fig:robustness}). The oscillation amplitudes for $10\, \Gamma$ are shown in the graphs. The model is stable for the nominal parameters (Table \ref{tab:parameters_plants}). Increasing robustness via $\Gamma$ lead to instability (Fig.\ \ref{fig:robustness} \textit{A}). During one oscillation period the internal zinc concentration reached up to 3.5 times the steady state value, meaning that strong and possible toxic periodic peaks of zinc are produced. These peaks exceed the steady state values of the less robust mechanism (Fig.\ \ref{fig:robustness} \textit{B}). Toxicity for high external zinc conditions could be either because of stable high internal zinc concentrations (non-robust mechanism) or due to toxic high amplitude oscillations (robust mechanism). Reducing robustness could be a strategy to avoid strong zinc bursts and cell might use other mechanisms to damp strong oscillations, such as buffering and sequestration.

\section*{Methods}

The ordinary differential equation systems were simulated with either an explicit eighth-order Runge-Kutta method or an implicit Rosenbrock stepper for stiff differential equations. Steady states were calculated by Newton's method in combination with a path following method for varying parameters. Jacobians were calculated analytically. The model parameters were determined by fitting the model to measurements. For this purpose, Brent's algorithm was applied to minimize $\chi^2$ \cite{Gegenfurtner_1992,Bevington_2003}. The standard deviation of a measurement was assumed to be proportional to its value and the relative error ($17\%$) was chosen such to obtain a reduced $\chi^2$ of the order of one. This way, low and high values had the same weights and were fitted equally well. Penalties were added to $\chi^2$ to avoid negative parameter values. The confidence intervals were obtained by calculation of the covariance matrix via the Hessian of $\chi^2$ \cite{Bevington_2003}. The measurements in \cite{Zhao_1998,Bird_2004} were combined and scaled correctly. Determination of part of the scaling factors were included into the fitting process, while the rest was prescribed with given values (personal communication of D. Eide).

\section*{Acknowledgments}

The authors thank Ansgar Bohman for his helpful comments and David Eide for supplying the yeast data scaling factors.\\

\bibliography{zincbib}

\clearpage
\section*{Figures}

\clearpage
\section*{Tables}

\begin{table}[!ht]
\centering
\begin{minipage}{3.27in}
\centering
\caption[Yeast: parameters]{\bf{Yeast: parameters}}
  \begin{tabular}{lrl}
  \hline
  Parameter & Value & $\pm$ s.d.\\
  \hline
  $K_{A}$ &  109 & $\pm38$\\
  $K_{1}$ & 450 & $\pm307$ \\
  $K_{2}$ & 444 & $\pm119$\\
  $K_{2}'$ & 2171 & $\pm1191$\\
  $\Gamma_A$ & 714 & $\pm600$ \\
  $\Gamma_{T1}$ & 29.6 & $\pm31.5$ \\
  $\kappa$ & 6.3 & $\pm3.0$\\
  $K^t_1$ / $\mu M$ & 139 & $\pm65$\\
  $K^t_2$ / $\mu M$ & 2584 & $\pm1511$\\
  \hline
  \end{tabular}
\begin{flushleft} Parameters values and standard deviations obtained by fitting the model to measurements published in \cite{Zhao_1998} and \cite{Bird_2004}.
\end{flushleft}
\label{tab:parameters_yeast}
\end{minipage}
\end{table}

\begin{table}
\centering
\begin{minipage}{3.27in}
\centering
\caption[Plant roots: parameters]{\bf{Plant roots: parameters}}
\begin{threeparttable}[p]
  \begin{tabular}{lrrrr}
  \hline
  Parameter & Act. only & Act. dimer. & \parbox[t]{2cm}{\centering Act./Inhib. dimer.}\\[5mm]
  \hline
  $K^t$ $[\mu M]$\tnote{*} & $13$ & $13$ & $13$\\  
  $K$ &  20 & 20 & 20\\
  $\Gamma_A$ & $41138$ & $1844$ & -- \\
  $\Gamma$ & -- & -- & $38$ \\
  $\Gamma'$ & -- & -- & $167.2$ \\
  $\Gamma_I$ & -- & -- & $1000$ \\
  $\zeta$\tnote{$\dagger$} & -- & -- & $4.4\cdot 10^{-3}$\\
  $\xi$\tnote{$\ddagger$} & -- & -- & $10^{-3}$\\
  \hline
  \end{tabular}
  \begin{tablenotes}\footnotesize 
  \item[] * Value for ZIP1, \cite{Grotz_1998};\quad $\dagger$ $\zeta= \Gamma' / \Gamma \Gamma_I$;\quad $\ddagger$ $\xi = 1/ \Gamma_I$.
  \end{tablenotes}
\end{threeparttable}
\begin{flushleft} Plant roots: parameters used in the simulation of the \emph{activator only}, \emph{dimerized activator} and the \emph{dimerized activator/inhibitor} models.
\end{flushleft}
\label{tab:parameters_plants}
\end{minipage}
\end{table}

\end{document}